\documentclass[sigconf,authorversion,preprint]{acmart}
\usepackage{caption}
\usepackage{listings}
\newlength\bibitemsep 

\usepackage{subfig}
\usepackage{hyperref}
\usepackage{tabularx}

\floatstyle{boxed}
\newfloat{definition}{thp}{lob}[section]
\floatname{definition}{Definition}

\setcopyright{rightsretained}

\newcommand{\TITLE}{The DEBS 2022 Grand Challenge: \\Detecting Trading Trends in Financial Tick Data}

\newcommand{\KEYWORDS}{Event processing, data streaming, trading, technical analysis}

\newcommand{\RM}{Ruben Mayer}
\newcommand{\RMEMAIL}{ruben.mayer@tum.de} 
\newcommand{\CD}{Christoph Doblander}
\newcommand{\CDEMAIL}{christoph.doblander@tum.de} 
\newcommand{\JT}{Jawad Tahir}
\newcommand{\JTEMAIL}{jawad.tahir@tum.de} 
\newcommand{\SF}{Sebastian Frischbier}
\newcommand{\SFEMAIL}{sebastian.frischbier@infrontfinance.com} 
\newcommand{\AH}{Arne Hormann}
\newcommand{\AHEMAIL}{arne.hormann@infrontquant.com} 
\newcommand{\HA}{Hans-Arno Jacobsen}
\newcommand{\HAEMAIL}{jacobsen@eecg.toronto.edu}

\newcommand{\RMADDR}{	\institution{Technical University of Munich}%
                        \country{Germany}}
                                      
\newcommand{\CDADDR}{	\institution{Technical University of Munich}%
                        \country{Germany}}   

\newcommand{\JTADDR}{	\institution{Technical University of Munich}%
                        \country{Germany}}

\newcommand{\SFADDR}{	\institution{Infront Financial Technology GmbH}%
                        \country{Germany}}

\newcommand{\AHADDR}{	\institution{Infront Quant AG}%
	\country{Germany}}

\newcommand{\HAADDR}{	\institution{University of Toronto}%
                        \country{Canada}}

\copyrightyear{2022}
\acmYear{2022}
\setcopyright{acmlicensed}\acmConference[DEBS '22]{The 16th ACM International Conference on Distributed and Event-based Systems}{June 27-June 30, 2022}{Copenhagen, Denmark}
\acmBooktitle{The 16th ACM International Conference on Distributed and Event-based Systems (DEBS '22), June 27-June 30, 2022, Copenhagen, Denmark}
\acmPrice{15.00}
\acmDOI{10.1145/3524860.3539645}
\acmISBN{978-1-4503-9308-9/22/06}

\begin{document}

\title{\TITLE}

\author{\SF} \affiliation{\SFADDR} \email{\SFEMAIL}\orcid{0000-0002-3517-8090}
\author{\JT} \affiliation{\JTADDR} \email{\JTEMAIL}\orcid{0000-0003-2008-7994}
\author{\CD} \affiliation{\CDADDR} \email{\CDEMAIL}
\author{\AH} \affiliation{\AHADDR} \email{\AHEMAIL}
\author{\RM} \affiliation{\RMADDR} \email{\RMEMAIL}\orcid{0000-0001-9870-7466}
\author{\HA} \affiliation{\HAADDR} \email{\HAEMAIL}\orcid{0000-0003-0813-0101}

\renewcommand{\shortauthors}{Frischbier et al.}

\begin{abstract}
The DEBS Grand Challenge (GC) is an annual programming compe\-tition open to practitioners from both academia and industry. The GC 2022 edition focuses on real-time complex event processing of high\--volume tick data provided by Infront Financial Technology GmbH. The goal of the challenge is to efficiently compute specific trend indicators and detect patterns in these indicators like those used by real-life traders to decide on buying or selling in financial markets. The data set \emph{Trading Data} used for benchmarking contains 289 million tick events from approximately 5500+ financial instruments that had been traded on the three major exchanges Amsterdam (NL), Paris (FR), and Frankfurt am Main (GER) over the course of a full week in 2021. The data set is made publicly available.
In addition to correctness and performance, submissions must explicitly focus on reusability and practicability. Hence, participants must address specific nonfunctional requirements and are asked to build upon open-source platforms.
This paper describes the required scenario and the data set \emph{Trading Data}, defines the queries of the problem state\-ment, and explains the enhancements made to the evaluation plat\-form \emph{Challenger}~that handles data distribution, dynamic sub\-scriptions, and remote eval\-ua\-tion of the submissions.

\end{abstract}



\maketitle

\section{Introduction}%
\label{sec:introduction}

The DEBS 2022 Grand Challenge (GC) is a programming challenge or\-ganized as part of the annual ACM International Conference on Distributed and Event-Based Systems (DEBS). The GC encourages participants from academia and industry to solve a practical problem by building an efficient and elegant distributed event-driven solution for it. Submitted solutions are benchmarked using real-world data and compared based on their performance, design, and practicability. 

This year's edition of the GC applies technical analysis to financial markets, featuring fine-granular real-world data about financial instruments such as equities and indices rarely available to research. The goal of this GC is to efficiently apply complex event processing at scale to detect opportunities in price movements hidden amongst millions of events. Participants need to implement trading strategies such as those used by real-life traders to decide on buying or selling of their assets. Consequently, multiple trend indicators must be tracked for each financial instrument's instance. Upon detecting specific patterns in these trend indicators, alerts need to be created if a trader has declared interest in this instrument instance; to raise the bar, our traders change their interest dynamically.

The remainder of this paper is structured as follows. In Section~\ref{sec:background}, we provide  background information on the domain of financial market data. In Section~\ref{sec:data}, we introduce the data set \emph{Trading Data} containing the real tick data used for this challenge. In Section~\ref{sec:query} we formalize the problem statement and in Section~\ref{sec:eval} we describe the evaluation done on the \emph{Challenger} platform~\cite{tahir10.1145} with the enhancements we added for this year's challenge. Finally, we conclude in Section~\ref{sec:conclusion}.


\section{Background}%
\label{sec:background}

Trading is fuelled by precise real-time event data together with reliable background information about financial instruments such as equities, indices or funds. Instances of a financial instrument are called \emph{symbols}. The high-volume streams of events reporting demand (\emph{ask}), supply (\emph{bid}), made trades (\emph{last}), and other information about each symbol are called financial \emph{market data}.  

The overall amount of market data pub\-lish\-ed by the various exchanges on a daily basis and  processed by technical solution providers such as Infront Financial Technology GmbH (formerly vwd Verei\-nigte Wirtschaftsdienste GmbH)  is massively increasing. For ex\-ample, the daily average number of events being processed by Infront increased from 18 billion in 2019 to 40 billion in 2021.

Leaving aside the special case of algorithmic trading, market data are provided to users at different levels of quality of information (QoI) depending on their subscription; quality attributes in this context refer primarily to granularity, timeliness, and completeness, ranging from fine-granular tick data to end-of-day aggregations.

\begin{figure}
	\centering
	\includegraphics[width=\linewidth]{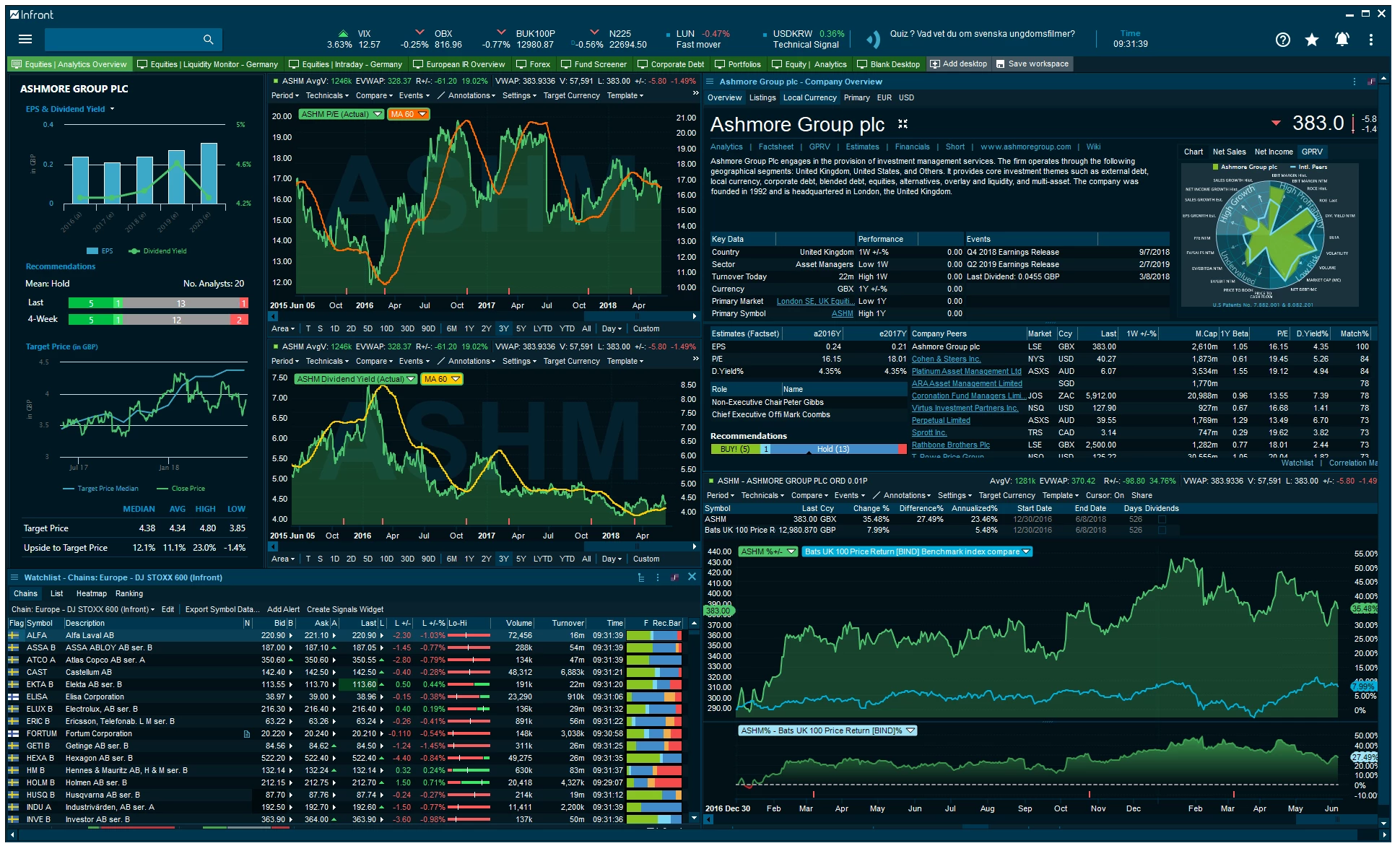}
    \caption{Example screen of a market data terminal product (here: Infront Professional Terminal).}
	\label{fig:marketdataterminalexample}
\end{figure}

Traders, analysts, and other stakeholders utilize market data of their required quality in interactive decision support systems called market data terminals to identify investment opportunities for specific sets of symbols they are interested in. The Infront Professional Terminal (IPT) shown in Fig.~\ref{fig:marketdataterminalexample} is an example of a terminal solution that fuses market data with metadata, news, interactive analytics, advanced visualizations, and direct trading functionality.

\section{Data Set: One Week of Real Tick Data}%
\label{sec:data}

The data used for this year's Grand Challenge are based on a week's worth of real tick data captured by Infront Financial Technology GmbH in 2021. The full data set intentionally contains far more data than is actually needed for the DEBS 2022 Grand Challenge to foster reuse in future research projects. The full data set \emph{Trading Data} used for the DEBS Grand Challenge 2022 is publicly available \cite{frischbier:debs2022GCTradingData} licensed under an open license\footnote{\url{http://creativecommons.org/licenses/by-nc-sa/4.0/}}. 

\subsection{Full Data Set: \emph{Trading Data} }

The full data set contains 289 million events consisting of tick data and housekeeping events recorded from Monday, November 8th, to Sunday, November 14th, 2021. The data cover 5504 equities and indices that are traded on three European exchanges (exchange code in parentheses): Paris (FR), Amsterdam (NL), and Frankfurt/Xetra (ETR). All tick data events for security type \emph{equities} (e.g., Royal Dutch Shell or Siemens Healthineers) and \emph{indices} are contained in the set as they had been captured on these days by Infront's systems. Consequently, this data set reflects real update rates per second and can also be used for complex event processing research tasks that need to correlate events for indices with those for equities. In the case of exchange-specific and ambiguous identifiers for the same symbol, these have already been normalized by Infront, while global date and CEST timestamp information have been added together with various metadata for initial enrichment. Note that the total order of events per symbol across exchanges cannot be assumed.

The distribution of events in the data set can be briefly summarized as follows: slicing by source, most traffic has been recorded from source Paris (55\%) followed by Amsterdam (28\%) and Frankfurt (17\%). Slicing by instrument type, we observe that more events are generated by activities regarding equities (239,5 million) than for indices (49,5 million). Actual trading activities (e.g., price events), however, are much higher for indices (49 million) than for equities (10 million). Furthermore, analysing the distribution of events across symbols reveals a long-tail distribution for general event types but also for trading activities regardless of exchange: most traffic is related to only a small number of symbols.

\begin{figure*}
	\centering
	\includegraphics[width=0.9\linewidth]{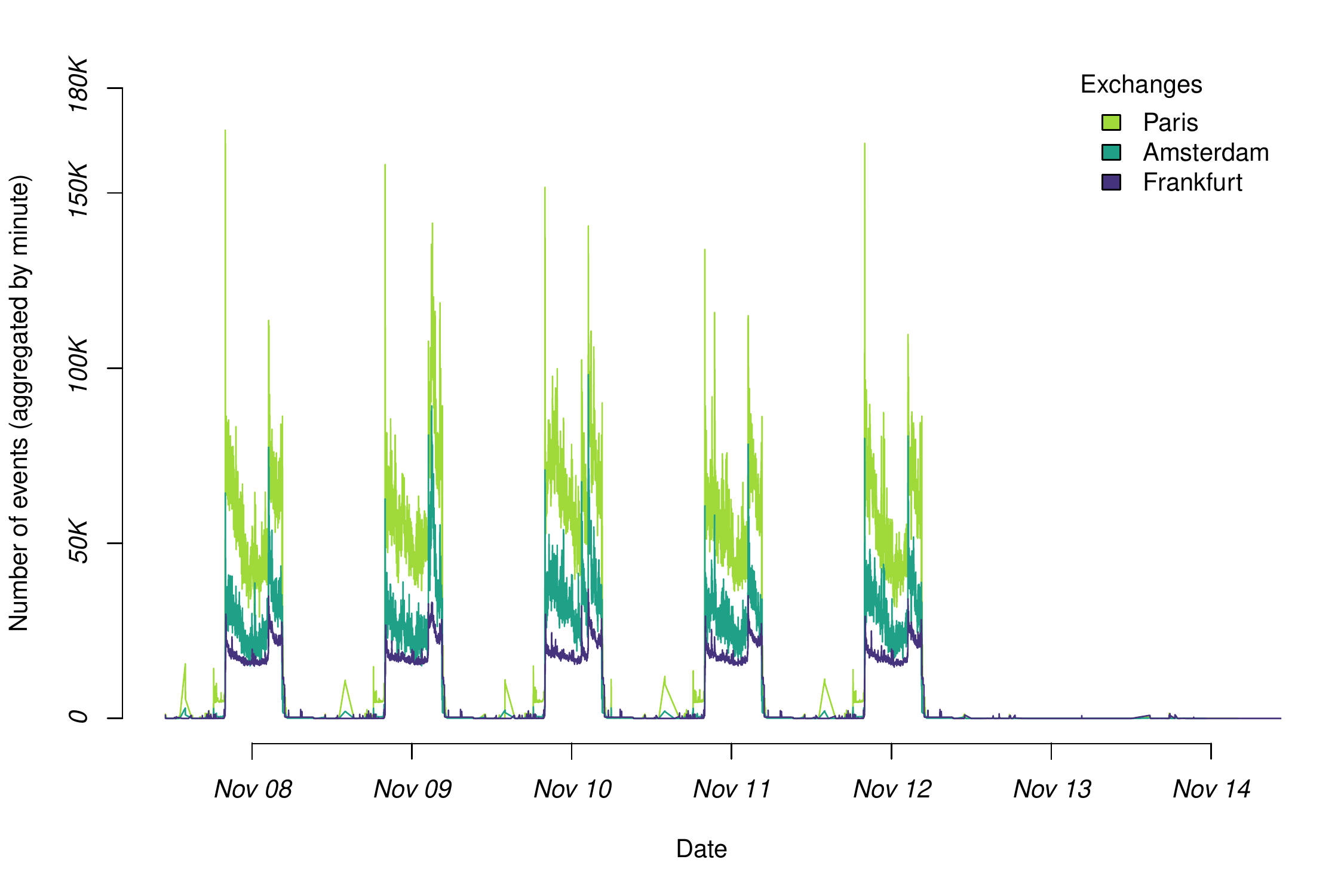}
    \caption{Daily load patterns in \emph{Trading Data} for all exchanges: start/end of trade, midday, and weekend.}
	\label{fig:gc-datasets-dailypatterns}
\end{figure*}

\begin{figure}
	\centering
	\includegraphics[width=\linewidth]{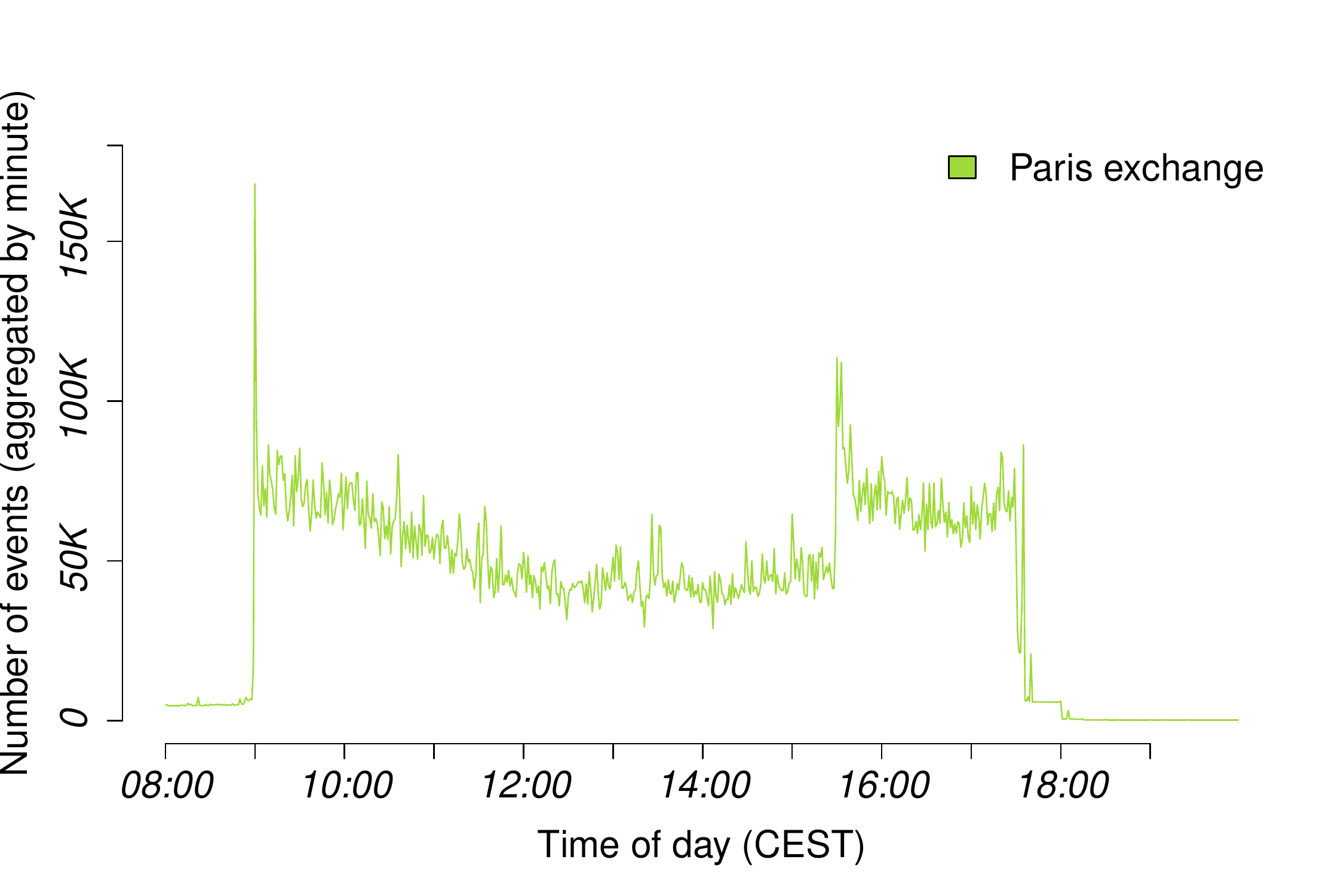}
   \caption{Paris exchange on Monday, November $8^{th}$.}
	\label{fig:gc-datasets-dailypatterns-detail}
\end{figure}

Plotting the cumulative update rates per exchange over time in Fig.~\ref{fig:gc-datasets-dailypatterns} shows the typical repetitive patterns per source. Zooming-in to a single trading day, as in Fig.~\ref{fig:gc-datasets-dailypatterns-detail} for the Paris exchange, shows the load spikes at opening and closing times while activity is lowest at local lunch time. Note that the spikes can only partially be attributed to direct trading activity but also to housekeeping tasks (e.g., maintaining order books, statistics) creating additional traffic. This pattern applies to an exchange in general as shown by Frischbier et al.~\cite{frischbier2019managing}.

\subsection{Subsets for GC 2022}
\label{subsec:gc-use}

\begin{figure}
	\centering
	\includegraphics[width=0.9\linewidth]{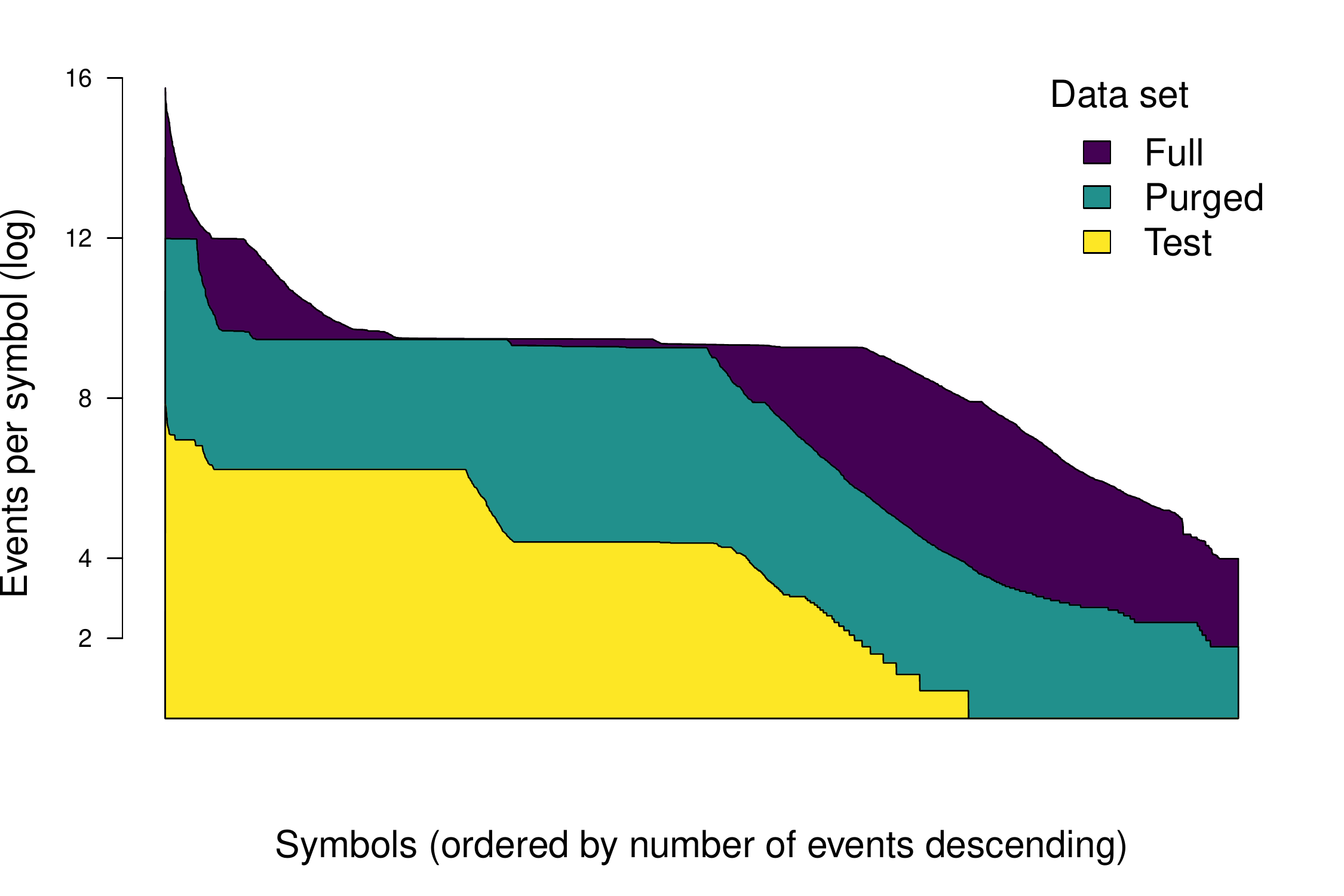}
    \caption{Long tail distribution of events in \emph{Trading Data}~and the two derived subsets \emph{Purged}~and \emph{Test}.}
	\label{fig:gc-datasets-distribution-symbols}
\end{figure}

For the GC 2022, two subsets, \emph{Purged} and \emph{Test}, were extracted from the full data set. The characteristic long-tail distribution of events per symbol is preserved in all data sets as shown in Fig.~\ref{fig:gc-datasets-distribution-symbols}.

\subsubsection{\emph{Purged} subset:} This data set contains only \emph{price} events relevant to the GC 2022 queries. This subset of 59 million events is used to evaluate the submitted solutions on the \emph{Challenger}~ platform. Consequently, participants do not need to filter out events that do not contain attributes relevant to the GC 2022. This data set contains price events for almost all symbols in the full data set (i.e., 5183/5504) with indices being more actively traded than equities (i.e., 82\% vs 18\%) and most traffic being recorded from source Frankfurt (54\%), followed by Paris (36\%) and Amsterdam (10\%).  

\subsubsection{\emph{Test} subset:} This data set contains the first one million events from the \emph{Purged}~data set for debugging and calibration by the participants when setting up their solution. Still covering 5177 symbols from both security types and all three sources but also preserving a long-tail distribution across symbols, it has been published upfront via \emph{Challenger} for registered participants as a representative sample. 

\subsection{Attributes and Format}%
\label{subsec:data-attributes}

For convenience and portability, Trading Data is provided as a collection of flat comma-separated values (CSV) files (one file per day). Each line in a file represents a single event. Event types (e.g., \emph{bid}, \emph{ask}, \emph{price}) are not explicitly marked but are identified by the respective non-NULL attributes (e.g., attribute \emph{ask} for an \emph{ask} event). 

The attributes available in events of the data sets are shown in Table~\ref{table:tradingdata-columns}; The attributes directly relevant for this GC are marked in the third column with a star ($\star$). Global CEST timestamps are in the format HH:MM:SS.ssss while dates are stored as DD-MM-YYYY.

\newcolumntype{L}{>{\raggedright\arraybackslash}X}
\begin{table}[!h]
	\centering
	\small
	\begin{tabularx}{\linewidth}{|l|l|L|c|} 
\hline
	\textbf{ID} & \textbf{Title} & \textbf{Description} & \textbf{GC}\\
\hline\hline
1  & ID.[Exchange] & Unique identifier for this symbol with trading exchange: Paris (FR) / Amsterdam (NL) / Frankfurt (ETR) & $\star$\\
2  & SecType 	  & Security type: [E]quity or [I]ndex & $\star$\\
3  & Date 		  & System date last received update &  \\
4  & Time		  & System time last received update  & \\
5  & Ask		  & Price of best ask order  & \\
6  & Ask volume	  & Volume of best ask order & \\
7  & Bid		  & Price of best bid order  & \\
8  & Bid volume	  & Volume of best bid order & \\
9  & Ask time	  & Time of last ask & \\
10 & Day's high ask & Day's high ask & \\
11 & Close 		  & Closing price (six digits) & \\
12 & Currency 	  & Currency (according to ISO 4217) & \\
13 & Day's high ask time & Day's high ask time & \\
14 & Day's high   & Day's high (price) & \\
15 & ISIN		  & ISIN (International Securities Identification Number) & \\
16 & Auction price & Price at midday's auction & \\
17 & Day's low ask & Lowest ask price of the current day & \\
18 & Day's low	  & Lowest price of the current day & \\
19 & Day's low ask time & Time of lowest ask price of the current day & \\
20 & Open		  & First price of current trading day & \\
21 & Nominal value & Nominal Value & \\
22 & Last		   & Last trade price & $\star$ \\
23 & Last volume   & Last trade volume & \\
24 & Trading time  & Time of last update (bid/ask/trade)  & $\star$\\
25 & Total volume  & Cumulative volume for current trading day & \\
26 & Mid price     & Mid price (between bid and ask) & \\
27 & Trading date  & Date of last trade & $\star$\\
28 & Profit		   & Profit & \\
29 & Current price & Current price & \\
30 & Related indices & Related indices  & \\
31 & Day high bid time & Time of day's highest bid & \\
32 & Day low bid time & Time of day's lowest bid & \\
33 & Open time & Time of open price  & \\
34 & Last price time & Time of last price & \\
35 & Close time & Time of closing price & \\
36 & Day high time & Time of day's high & \\
37 & Day low time & Time of day's low & \\
38 & Bid time & Time of last bid update & \\
39 & Auction time & Time of last auction price & \\
\hline	
\end{tabularx}
\caption{Attributes in \emph{Trading Data}: syntax and semantics.}
\label{table:tradingdata-columns}
\end{table}

Some events appear to come with no payload in the full data set. We had to balance the desire for a raw and unabridged data set with the need to protect intellectual property and to keep the data set's size as small as possible. In particular, only a small subset of attributes needs to be evaluated in this year's GC. Hence, we eliminated certain attributes to preserve the number of events and their update patterns over time while reducing the overall size.

\section{Problem Definition}%
\label{sec:query}

This year’s GC requires participants to imple\-ment a basic trading strategy as applied by intraday traders in real life. The strategy aims at (a) identifying trends in price movements for individual symbols using event aggregation over tumbling windows (Query 1) and (b) generating buy/sell advisories  upon detecting specific patterns using complex event processing (Query 2). 
 
\subsection{Definitions and Relaxations}%
\label{sec:relaxations}

We define the following terms and relaxations to allow participants to focus on the actual engineering aspects of the GC:

\begin{enumerate}
	\item Each instrument instance is identified by a \emph{symbol} $s \in S$ consisting of a unique string and the exchange code of the exchange the instrument instance is being traded on;
	e.g., the symbol \emph{RDSA.NL} denotes shares of \emph{Royal Dutch Shell} traded on the Amsterdam exchange\footnote{\url{https://live.euronext.com/product/equities/GB00B03MLX29-XAMS/quotes}}.
	\item All events have to be grouped by their symbol.
	\item All events provided via \emph{Challenger}~are prefiltered to contain only pricing information per symbol $s$.
	\item All metrics for all $s \in S$ must constantly be calculated, but updates need to be provided only for a subset of symbols $s \in \overline{S} \subsetneq S$ the platform subscribes to (simulating traders).
	\item Calculations are based on windows of 5 minutes’ length.
	\item The first window $w_0$ starts at 0:00 (midnight) CEST.
	\item Windows do not overlap (tumbling windows).
	\item A window $w_i$ is evaluated once the next window $w_{i+1}$ starts.
\end{enumerate}

\subsection{Quantitative Indicators (Query 1)} 

The first query defines one of the most essential indicators for each symbol used in technical analysis to identify trends: the \emph{exponential moving average} (EMA). Multiple price events observed within $w$ minutes define a window of length $w$ (e.g., $w = 5$ minutes). In our example a window of length $w$ cannot be evaluated until the next window starts. The EMA of the current window is calculated by weighting the price last observed in the current window with the EMA of the previous window. Furthermore, we define $EMA^j_{s,0}=0$, i.e., for the first reading of a symbol $s$  in the very first window, we assume the EMA of the previous window is zero.

\begin{equation*}
{EMA}^j_{s,w_{i}} = \bigg[{Close}_{s,w_i} \cdot \bigg(\frac{2}{1+j}\bigg)\bigg] + \underbrace{{EMA}^j_{s,w_{i-1}}}_{\text{prev. window}} \cdot \bigg[1-\bigg(\frac{2}{1+j}\bigg)\bigg]
\end{equation*}

with

\begin{eqnarray*}
|w| & : & \text{window duration in minutes}\\
j & : & \text{smoothing factor for EMA with } j \in \{38, 100\}\\
s & : & \text{symbol } s \in S = \{s_1,\dots,s_n\} \text{}\\
{Close_{s,w_i}} & : & \text{last price event for } s \text { observed in  window } w_i\\
{EMA}^j_{s,w_{0}} & = & 0
\end{eqnarray*}

\subsection{Breakout Patterns: Crossovers (Query 2)}

The quantitative indicators of Query 1 are used in Query 2: \emph{breakout patterns} can be identified by tracking two EMAs per symbol that are computed over different intervals. 

Generally, breakout patterns describe meaningful changes in the development of a price that indicate the start of a \emph{trend} (even if only temporary). A change is called a \emph{bullish breakout}, if the price is starting to rise steadily (crossover from below / breaking through the \emph{support area}) and a \emph{bearish breakout} if the price is going to lose steadily (crossover from above / breaking through the \emph{resistance area}). 

Properly identifying such changes and their nature in a timely manner allows a trader to monetize this knowledge by immediately buying (in case of a bullish breakout) or selling (in case of a bearish breakout) to maximize revenue.

\subsubsection{Bullish Breakout Pattern: Buy Advisory.}

Generally, we detect a bullish breakout pattern for a symbol once the EMA with the shorter interval $j_1$ starts to overtake the EMA with the longer interval $j_2$. In this case, a \emph{buy advise event} must be created immediately so that a trader can still benefit from a relatively low price. 

For long intervals of $j_1=50$ days and $j_2=100$ days this crossover is specifically called a \emph{golden cross} to indicate a golden opportunity for long-term investments. 

\begin{figure}[!ht]
	\centering
   \includegraphics[width=0.8\linewidth]{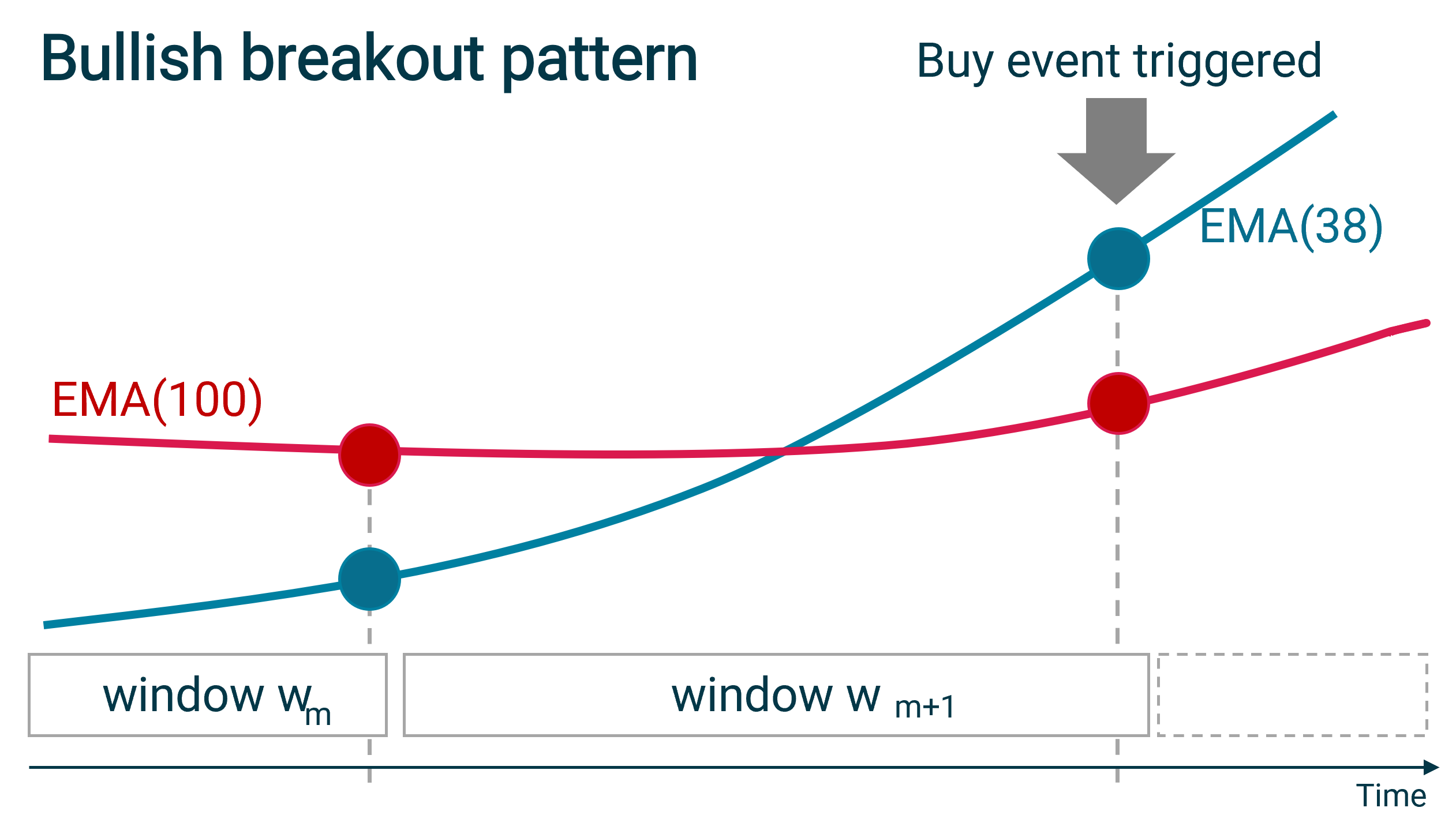}
    \caption{Crossover called a \emph{bullish} breakout pattern.}
    \label{fig:bullish-breakout}
\end{figure}

For this challenge, we use a granularity of minutes by setting $j_1=38$ and $j_2=100$ and create a \emph{buy advise event} upon detecting a crossover as illustrated in Fig.~\ref{fig:bullish-breakout} and formalised in Equations~\ref{eqn:bullish-breakout1} and ~\ref{eqn:bullish-breakout2}.

A bullish breakout pattern can be observed if and only if
\begin{eqnarray}
		{EMA}^{38}_{s,w_i} & > & {EMA}^{100}_{s,w_i} \text{ and}\label{eqn:bullish-breakout1}\\
		{EMA}^{38}_{s,w_{i-1}} & \le & {EMA}^{100}_{s,w_{i-1}} \label{eqn:bullish-breakout2}
\end{eqnarray}

	Subsequently, a \emph{buy advise} event must be generated.

\subsubsection{Bearish Breakout Pattern: Sell Advisory.}

Generally, we detect a bearish breakout pattern for a symbol once the EMA with the longer interval $j_2$ starts to overtake the EMA with the shorter interval $j_1$. In this case, a \emph{sell advise event} must be created immediately so that a trader can still sell at a relatively high price.  

Conversely to the golden cross described earlier, a bearish pattern for $j_1$=50 days and $j_2$=100 days is specifically called a \emph{death cross}. 

\begin{figure}[!ht]
	\centering
	\includegraphics[width=0.8\linewidth]{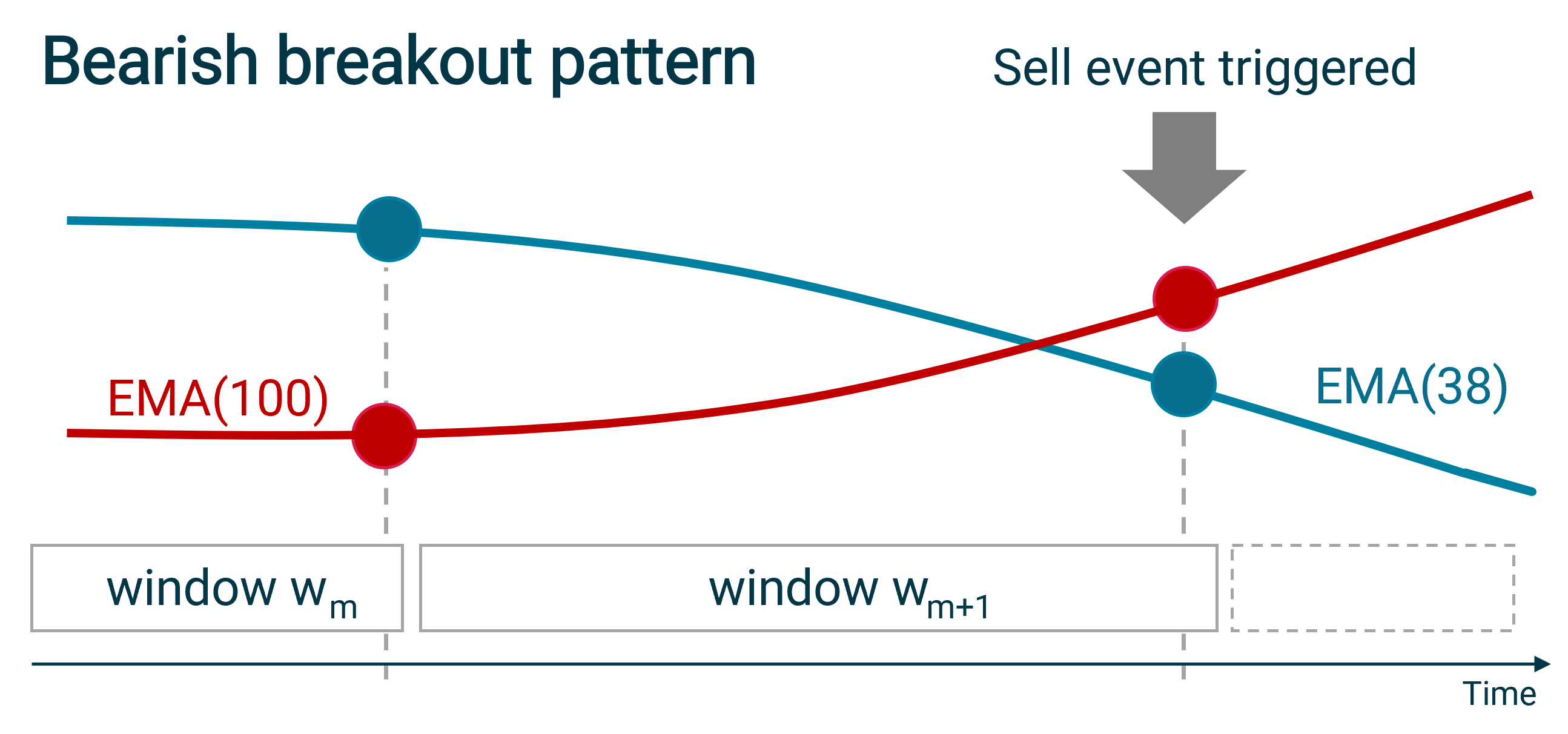}
    \caption{Crossover called a \emph{bearish} breakout pattern}
    \label{fig:bearish-breakout}
\end{figure}

For the bullish pattern, we use a granularity of minutes by setting $j_1=38$ and $j_2=100$ and create a sell advise event as shown in Fig.~\ref{fig:bearish-breakout} and formalised in Equations~\ref{eqn:bearish-breakout1} and~\ref{eqn:bearish-breakout2}.

A bearish breakout pattern can be observed if and only if\\

\begin{eqnarray}
	{EMA}^{38}_{s,w_i} & < & {EMA}^{100}_{s,w_i} \text{ and}\label{eqn:bearish-breakout1} \\
	{EMA}^{38}_{s,w_{i-1}} & \ge & {EMA}^{100}_{s,w_{i-1}}\label{eqn:bearish-breakout2}
\end{eqnarray}

Subsequently, a \emph{sell advice event} must be generated.

\subsection{Bonus: Smart Visualization}

Whether users can fully exploit a decision support system for trading does not only depend on the correctness and performance of its implementation; being able to visually cut through the noise and visually emphasize the relevant data to the user is almost as important. Hence, GC participants were encouraged to find a smart way to visualize the results of the queries for bonus points.

\section{Evaluation}%
\label{sec:eval}
 
Submitted solutions are automatically evaluated by the distributed evaluation platform \emph{Challenger}~\cite{tahir10.1145} and ranked by their performance regarding latency, processing time, and throughput. Furthermore, each solution must address specific nonfunctional requirements to prove its practicability and portability beyond the boundaries of this community and GC scenario. 
In this section, we describe the automatic performance evaluation performed by our evaluation platform, the enhance\-ments we added to the \emph{Challenger}~ platform\footnote{\url{https://github.com/jawadtahir/CHALLENGER}} to this end, and briefly review evaluation guidelines that aim at engendering reusability in the solutions.

 \subsection{Evaluation Approach}
 
The functional evaluation of the solutions covers two aspects, namely correctness and performance. These are centrally gauged using \emph{Challenger} in the same manner as in the previous GC edition~\cite{tahir10.1145}.
The correct\-ness of a solution is assessed by code reviews. 
For performance evaluation, the latency ($90^{th}$ percentile) of the queries is averaged per solution and ranked among all received solutions. The team with the lowest rank wins the challenge.

 \subsection{Expected Results in the Evaluation}%

For a fair evaluation, our platform mimics the usual behaviour of traders using market terminal solutions: traders subscribe to individual sets of symbols they want to track and get informed about opportunities (e.g., advise to buy or sell).

Every batch of events pushed to a participant's solution comes with a list of symbols that the \emph{Challenger}~ platform subscribes to in order to simulate a trader. The subscription is valid until it is updated (i.e., the current set of symbols is replaced with a new one). Subscriptions change dynamically over the evaluation session. Consequently, participants always need to keep track of all symbols with their EMAs and crossovers to properly reply to the latest subscriptions. Subscription patterns are implemented to be unpredictable but reproducible.

For each symbol $s \in \overline{S}$ that \emph{Challenger}~ has currently been subscribed to, the evaluation platform expects the following results returned from a submitted solution per batch: 

\begin{itemize}
    \item Query 1: ${EMA}^{38}_{s,w_{last}}$ and ${EMA}^{100}_{s,w_{last}}$
    \item Query 2: last three \emph{sell/buy advice events}
\end{itemize}

\subsection{Evaluation Platform}%
\label{sec:platform}

For competitive benchmarking of the submitted solutions, we have reused and extended the evaluation platform \emph{Challenger}~\cite{tahir10.1145} that had been introduced for the ACM DEBS 2021 Grand Challenge.

\paragraph{About the platform.} \emph{Challenger}\ is a gRPC\footnote{https://grpc.io}-based service that allows participants to remotely benchmark their solutions with different but centrally provided data sets. 
The platform provides registered participants with dedicated virtual machines for running benchmarks. Solutions can be implemented with the programming language preferred by the participants using service stubs generated from the gRPC interfaces. \emph{Challenger} also provides participants with dashboards about their solution's performance and their ranking among all participants; data from these dashboards can be exported.

\begin{figure}[!ht]
	\centering
   \includegraphics[width=0.9\linewidth]{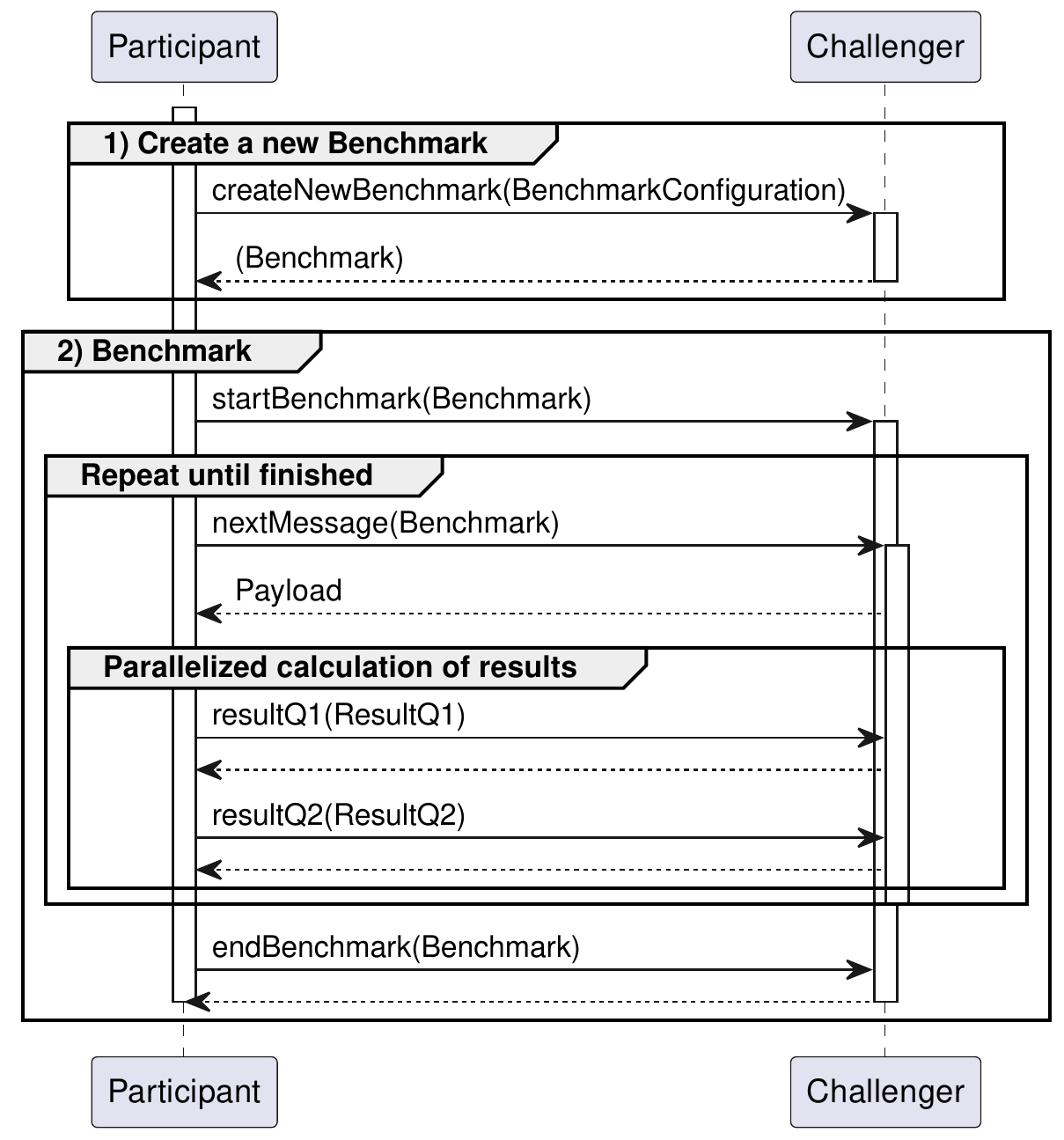}
    \caption{Challenger API.}
    \label{fig:challenger-api}
\end{figure}	

\paragraph{How to use the platform.} Participants generate stubs to communicate with the eva\-lu\-a\-tion platform using a client generator for gRPC. In the first step, par\-ti\-ci\-pants create a new benchmark; see ~Fig.~\ref{fig:challenger-api} (step~1). Then they invoke \emph{startBenchmark}, see~Fig.~\ref{fig:challenger-api} (step~2), which sets the starting point of the evaluation run on \emph{Challenger}. Next, the participants retrieve all batches and return two types of results for each batch. Timestamps in nanosecond resolution are taken from each retrieval of a batch and from each submission of the results. When the last batch is received, which is marked by a flag, the participants call \emph{endBenchmark}. The throughput for each query is calculated from the number of batches received between the start and end of the benchmark. The processing latency of each query is derived from the timestamps of the retrieval and submission of the results.

\begin{definition}
\small
\begin{lstlisting}
while True:
    batch = stub.nextBatch(benchmark)
    event_count = event_count +
                    len(batch.events)

    def queryResults(symbols:list[str]) 
        -> list[ch.Indicator]:
        
        # Your part: calculate the indicators 
        # for the given symbols
        
        return list()

    resultQ1 = ch.ResultQ1(
        #The id of the benchmark
        benchmark_id=benchmark.id,
        
        #The id of the benchmark
        batch_seq_id=batch.seq_id,
        indicators=queryResults(
            batch.lookup_symbols))
    
    # send the result of query 1 back
    stub.resultQ1(resultQ1)  
    
    def crossoverEvents() ->
        list[ch.CrossoverEvent]:
    
        #Your part: calculate the crossover events
    
        return list()

    # do the same for Q2
    resultQ2 = ch.ResultQ2(
        #The id of the benchmark
        benchmark_id=benchmark.id,
        
        #The sequence id of the batch
        batch_seq_id=batch.seq_id, 
        crossover_events=crossoverEvents()) 
    
    # submit the results of Q2
    stub.resultQ2(resultQ2) 
    
    # Step 4 - once the last event 
    # is received, stop the clock
    # See the statistics within ~5min here:
    # https://challenge.msrg.in.tum.de/benchmarks/
    
    if batch.last:
        print(f"received last batch, 
            total batches: {event_count}")
        stub.endBenchmark(benchmark)
        break
        

\end{lstlisting}    
\caption{Example code snippet in Python provided to participants about interacting with \emph{Challenger}.}
\label{lst:challenger-example}
\end{definition}

\subsection{Nonfunctional Requirements}
\label{sec:NFR}

The ACM DEBS Grand Challenge's primary goal is to complement conceptual research in the area of event-driven and distributed systems with the engineering challenge of putting such approaches into practice by solving problems using real-world data.  

Over the course of the years, participants focused more on the correctness and performance of their implementation than on maximizing the reusability and versatility of their solution beyond the scope of the challenge at hand. Understandably, most submitted solutions thus tended to be built from scratch and were custom-tailored to a single-node architecture to side-step the challenges that come with distributed systems. 

Previous instances of the GC already tried to motivate parti\-ci\-pants to consider nonfunctional requirements (NFRs) in their submission \cite{debsgc2014,gulisano2020debs}. Building on and reinforcing this trend, we introduced detailed NFRs in last year's edition of the GC and made them mandatory for this year's edition. With these we want to ensure a minimum level of portability and practicability beyond the scope of the current challenge. Hard NFRs to be explicitly addressed by participants are: configurability, scalability (with horizontal scalability being preferred), operational reliability/resilience, accessibility of the solution's source code, integration with standard tools and protocols, as well as documentation.

Most of these NFRs can directly be addressed by building upon widely-used industry-strength open-source platforms such as those curated by recognized open source foundations\footnote{\url{https://opensource.com/resources/organizations}}. We are very glad to report that this year all submitted solutions are built on top of various open-source distributed frameworks such as Apache Flink, Apache Kafka, Apache Spark, and Jupyter Notebook.

\section{Conclusions}
\label{sec:conclusion}

Registration for the 2022 edition of the GC opened in December 2021. Participants were required to first register on Microsoft CMT for general communication and the final submission of their short papers to the peer-review process. Successfully registered participants were then activated on the \emph{Challenger} platform. 
A total of 17 teams from academic and industrial organizations registered for GC 2022 and actively used \emph{Challenger} to benchmark their solutions against the \emph{Trading Data} set. By the deadline, eight teams were able to submit solutions to the peer-review process.

\begin{acks}
The authors would like to thank Christian Roth, Thorsten Hammel, Alexander Echler, and Fredrik Koch of the Infront group for their input and support. The infrastructure for running \emph{Challenger} was provided by the Chair for Application and Middleware Systems at Technical University of Munich (TUM).
\end{acks}


\bibliographystyle{ACM-Reference-Format}
\bibliography{debs2022gc}

\end{document}